# Title: Large Quantum Anomalous Hall Effect in Spin-Orbit Proximitized Rhombohedral Graphene

**Authors:** Tonghang Han[1]†, Zhengguang Lu[1]†, Yuxuan Yao[1]†, Jixiang Yang[1], Junseok Seo[1], Chiho Yoon[2], Kenji Watanabe[3], Takashi Taniguchi[4], Liang Fu[1], Fan Zhang[2], Long Ju[1]*

**Affiliations:**

[1]Department of Physics, Massachusetts Institute of Technology, Cambridge, MA, USA.

[2]Department of Physics, The University of Texas at Dallas, Richardson, TX, USA

[3]Research Center for Electronic and Optical Materials, National Institute for Materials Science, 1-1 Namiki, Tsukuba 305-0044, Japan

[4]Research Center for Materials Nanoarchitectonics, National Institute for Materials Science, 1-1 Namiki, Tsukuba 305-0044, Japan

*Corresponding author. Email: longju@mit.edu †These authors contributed equally to this work.

**Abstract:** The quantum anomalous Hall effect (QAHE) is a robust topological phenomenon featuring quantized Hall resistance at zero magnetic field. We report the QAHE in a rhombohedral pentalayer graphene/monolayer $WS_2$ heterostructure. Distinct from other experimentally confirmed QAHE systems, this system has neither magnetic element nor moiré superlattice effect. The QAH states emerge at charge neutrality and feature Chern numbers $C = \pm 5$ at temperatures up to about 1.5 K. This large QAHE arises from the synergy of the electron correlation in intrinsic flat bands of pentalayer graphene, the gate-tuning effect, and the proximity-induced Ising spin-orbit-coupling. Our experiment demonstrates the potential of crystalline two-dimensional materials for intertwined electron correlation and band topology physics, and may enable a route for engineering chiral Majorana edge states.

**Main Text:**

In 1988, Haldane proposed a prototype model to realize quantized Hall conductance in the absence of a magnetic field (*1*). It has informed the theoretical and experimental exploration of topological phases of matter for more than three decades. In the Haldane model, the graphene K and K' valleys carry opposite signs of mass (band gap), which amounts to the same sign of Berry curvatures that add up constructively to result in a Chern number of 1. The physical realization of the Haldane mass is very challenging, however, owing to the required complex hoppings that break chiral and time-reversal-symmetries at the atomic scale. Although the Berry curvatures of the Haldane model have been demonstrated in a cold-atom experiment (*2*), QAHE as the most prominent macroscopic signature of the Haldane model could not be measured there. In solid-state systems, the QAHE has been observed in two categories of materials (*3*): 1. Magnetic topological insulators such as Cr-/V-doped $(Bi,Sb)_2Te_3$ (*4–6*) and intrinsic $MnBi_2Te_4$ (*7*), where the time-reversal-symmetry is broken by the ordering of magnetic elements and the Berry curvatures of the two $\Gamma$ valleys of opposite surfaces add up to a Chern number of 1; 2. 2D materials with moiré superlattice at non-zero charge densities that correspond to odd filling factors, where the time-reversal-symmetry is broken spontaneously by the exchange interactions and the Chern number arises from a single principal valley (*8–14*). Without stacking multiple (effectively decoupled) periods of molecular-beam-epitaxy-grown quantum wells in the vertical direction (*15*), the largest Chern number that has been realized so far is 2 in ferromagnetic systems which were not fully quantized at zero magnetic field (*16, 17*). Is it possible to realize QAHE without magnetic elements or moiré engineering? Can we get a larger Chern number in other QAHE systems? Such possibilities have remained elusive experimentally, although integer QAHE in rhombohedral graphene was predicted to be possible by theory (*18*).

Crystalline graphene in the rhombohedral stacking order has recently been demonstrated as a platform for rich correlated and topological phenomena (*18–30*), without the need of a moiré superlattice. At non-zero charge carrier densities in the gate-induced and naturally existing flat bands, various isospin-symmetry-broken states and (unconventional) superconductivities have been observed (*22–26*). Near zero density, both tetralayer and pentalayer graphene showed a layer-anti-ferromagnetic state (LAF) driven by the large density-of-states and large pseudospin winding numbers in the pair of touching flat bands at zero energy (*24, 26, 27*). In the LAF state, electrons with opposite spins polarize to the top and bottom layers of graphene respectively, and the degeneracy of spin-polarized bands can be broken by applying a vertical gate-displacement field *D* (*18, 30*). Tuning through a critical value of *D*, the bandgap corresponding to one spin species can be gradually inverted, resulting in an insulator-metal-insulator transition. Although Chern insulator states can be induced in pentalayer graphene when the valley-degeneracy is broken explicitly by a perpendicular magnetic field (*24*), a QAH state at zero charge density and zero magnetic field was absent. As an alternative, the spin-orbit-coupling (SOC) effect in graphene proximitized by transition metal dichalcogenides has been theoretically calculated, and experimentally demonstrated to be a powerful approach to modulate electron correlations in graphene-based systems (*31–41*). The valley-contrasting feature of such a proximity-induced SOC effect provides a powerful knob to break the valley-degeneracy in the LAF state—offering a promising approach to realize QAHE in crystalline rhombohedral multilayer graphene.

Here we report the QAHE in pentalayer rhombohedral graphene proximitized by a monolayer $WS_2$. This system is free of magnetic elements or moiré effects (at relevant energy scales)—distinct from previous systems that host the QAHE. We observed quantized Hall resistance $R_{xy} = \pm\frac{h}{5e^2}$ at zero magnetic field and zero charge density, where *h* is the Planck constant and $e$ is the electron charge. These states correspond to Chern numbers $C = \pm 5$, which is the largest (together with an engineered vertical superlattice of magnetic topological insulator (*15*)) among all experimentally realized QAHE systems. The quantized $R_{xy}$ persists to a temperature of ~1.5 K and shows clear magnetic hysteresis. By continuously tuning $D$, we observed topological phase transitions between topologically trivial and non-trivial states, accompanied by a magnetic phase transition from the LAF state to an orbital ferromagnetic state. We interpret the transitions and the QAH states using a minimal model that incorporates the LAF state, the gate displacement field and the proximity-induced Ising SOC.

## Phase Diagram of Rhombohedral Pentalayer Graphene/Monolayer WS$_2$

The device structure is shown in Fig. 1A, where the WS$_2$ layer is in proximity with the top layer of pentalayer graphene. We purposely chose a large twist angle between graphene and the bottom hBN to avoid the moiré effect (*16, 42–44*). The twist angle between WS$_2$ and graphene is estimated to be ~7 degrees (*45*). The device is fabricated into dual-graphite-gated Hall bars (Fig. 1B), where the combinations of the top and bottom gate voltages control the electron density $n_e$ and $D$ independently (*45*). We chose WS$_2$ because of its stronger SOC effect on graphene than that of WSe$_2$ predicted by a previous calculation (*34*).

Figure 1C shows the $n_e$-$D$ map of the longitudinal resistance $R_{xx}$ at $B = 0$ T and nominal fridge base temperature $T = 20$ mK. Compared with pentalayer graphene without a neighboring WS$_2$, most of the features are similar and we labeled the three characteristic states as LAF, LP (Layer-Polarized band insulator) and Bubble (*24, 25*). However, a dramatic difference is induced by WS$_2$ in the region outlined by the dashed square, at the center of which $R_{xx}$ shows values close to zero. Figures 1D and E show the zoomed-in maps of $R_{xx}$ and the Hall resistance $R_{xy}$ corresponding to the dashed box in Fig. 1C. $R_{xx}$ is symmetrized and $R_{xy}$ is anti-symmetrized with respect to a small magnetic field $B = \pm 0.1$ T to stabilize the magnetic polarization. In a rectangle region centered at $n_e = 0$ cm$^{-2}$ and $D/\varepsilon_0 = -0.165$ V/nm, $R_{xx}$ shows values below 100 Ω whereas $R_{xy}$ shows values of ~ 5 kΩ. Given the small value of magnetic field applied, the large value of $R_{xy}$ that does not change sign with charge carrier type is a strong evidence of anomalous Hall resistance. Such anomalous behaviors of $R_{xx}$ and $R_{xy}$ are absent on the positive $D$ side (corresponding data are shown in Fig. S2 (*45*)).

**QAHE with Chern Number $C = \pm 5$**

Next, we examine the anomalous Hall effect in detail. Figures 2, A and B, show a clear hysteresis of $R_{xy}$ and $R_{xx}$ at $n_e = 0$ cm$^{-2}$, $D/\varepsilon_0 = -0.165$ V/nm and temperature $T = 0.2$-4.5 K, when an out-of-plane magnetic field $B$ is swept back and forth. At $B = 0$ T, $R_{xy}$ is quantized at $\pm\frac{h}{5e^2}$(precise within the noise level) and $R_{xx}$ is less than 100 Ω at the lowest temperature. These key observations indicate QAH insulators with Chern numbers $C = \pm 5$. The QAH states persists to ~1.5 K, before the anomalous $R_{xy}$ starts to deviate from the quantized value, as shown in Fig. 2C. The QAH states occupy a range of electron density $\Delta n_e \sim 0.5*10^{11}$ cm$^{-2}$, as shown by the plateaus of $R_{xy}$ in Fig. 2D. In addition, the Landau fan measurements of $R_{xx}$ and $R_{xy}$ at the base temperature at $D/\varepsilon_0 = -0.165$ V/nm (Fig. 2, E and F) show evidence consistent with the $C = \pm 5$ QAH insulators. Firstly, the slope of $n_e$ vs $B$ aligns well with $C = \pm 5$ (indicated by the dashed lines) according to the Streda's formula (*46*). Secondly, these states persist to $B = 0$ T, whereas features corresponding to integer quantum Hall states appear only at above 0.5 T on the electron side.

Phenomenologically, the QAHE we have observed is distinct from those reported in the literature in three aspects. Firstly, the time-reversal-symmetry is broken by intrinsic orbital magnetism, in contrast to being broken by the ordering of magnetic elements in Cr-/V-doped $Bi_2Te_3$ and $MnBi_2Te_4$ (*4–7*). Secondly, the QAH states in pentalayer graphene/$WS_2$ reside at zero charge density, in contrast to non-zero charge densities and filling factors of a superlattice of all moiré-based 2D material devices (*8–14*). Thirdly, the Chern numbers of ±5 are larger than those in other experimentally demonstrated QAHE systems, with the exception of the largest Chern number in a vertically stacked superlattice of magnetically doped topological insulator (*15*). We will discuss the origins and implications of these differences in later sections.

**Topological Phase Transitions Driven by *D***

With the QAH states established, we further study the evolution of the ground state driven by the gate displacement field *D*. Figures 3A and B, show the temperature dependence of $R_{xx}$ and $R_{xy}$ as a function of *D* at $n_e = 0$ cm$^{-2}$. The device exhibits distinct behaviors in three regions: I. At $D/\varepsilon_0 <$ -0.22 V/nm and $D/\varepsilon_0 >$ -0.12 V/nm, $R_{xx}$ decreases as *T* increases whereas $R_{xy}$ fluctuates around zero (the fluctuations are caused by the large $R_{xx}$ values at low temperatures) — as expected for topologically trivial insulators; II. At $D/\varepsilon_0 \in$ (-0.22, -0.18) V/nm and $D/\varepsilon_0 \in$ (-0.15, -0.12) V/nm, $R_{xx}$ increases as *T* increases whereas $R_{xy}$ remains almost zero; III. At $D/\varepsilon_0 \in$ (-0.18, -0.15) V/nm, $R_{xx}$ increases as *T* increases whereas $R_{xy}$ decreases — as expected for a QAH insulator. Figure 3C shows further magneto-resistance measurements at the base temperature at $D/\varepsilon_0 \in$ (-0.22, -0.18) V/nm and $D/\varepsilon_0 \in$ (-0.15, -0.12) V/nm, which follows a $B^2$-dependence. Together with the fact that $n_e = 0$ cm$^{-2}$, the ground state in region II could be a compensated semi-metal. The gap size of the QAH state can be extracted from the temperature-dependence of $R_{xx}$ and $\delta R_{xy} = \frac{h}{5e^2} - R_{xy}$ as shown in Fig. 3D. Both fittings show a gap size of ~20 K, which is an order of magnitude larger than the 1.5 K and possibly implies the correlated nature of the QAH states. Figure 3E summarizes the band gap (extracted based on the thermal activation model for the LP and LAF states, see Fig. S4 for details (*45*)) as a function of *D*, which clearly shows continuously diminishing gaps when *D* approaches the semi-metal regions.

The data shown in Fig. 3 reveal topological phase transitions driven by *D*, where the electronic band structure changes between topologically trivial and non-trivial through gap inversion. We note that the peak of $R_{xy}$ shifts from $D/\varepsilon_0$ = -0.165 V/nm towards zero as the temperature is increased, as indicated by the green dashed curve in Fig. 3B. This is mainly because the electron correlation gap in the LAF state weakens and the *D* range of LAF reduces as the temperature is increased (*24*) (see Fig. S3 for more details (*45*)) — driving the phase boundary between LAF and QAH towards smaller absolute values of *D*.

## QAHE based on the Haldane Mechanism

There has not been a direct theoretical proposal to realize the QAHE (*18*) with the recipe demonstrated in our experiment. Based on all our experimental observations, we provide a minimal model to elucidate the topological phase transitions and the QAH states by using all the ingredients one by one. 1. In the LAF state (*18, 30*) at zero *D* and zero $n_e$, the spin-up and spin-down bands open up quasiparticle bandgaps of the same amplitude $\Delta_{\text{LAF}}$, but opposite in sign — meaning the spin-up (labeled by ↑) occupied states near the valence band top polarize to the top layer, whereas the spin-down (labeled by ↓) occupied states near the valence band top polarize to the bottom layer, as shown in Fig. 4A. Each isospin copy carries a Chern number as $C = \frac{5}{2}\sigma_z\tau_z$, where $\sigma_z$ and $\tau_z$ corresponds to the spin and valley index (±1), respectively (*18*). The Chern numbers of all four isospins sum up to zero and result in a topologically trivial state. 2. By applying a non-zero *D*, the gaps in spin-up bands shrink whereas those in the spin-down bands enlarge, because the specific

*D* direction prefers the bottom-layer polarization. We can now focus on the spin-up bands as they are closer to band inversion that is crucial to the topological properties of the whole system. This fact essentially reduces the model to a spinless model, as is the case for the Haldane model. 3. Introducing the Ising SOC (*31–35*) $\lambda_I \tau_z \sigma_z$ at the top layer further breaks the valley-symmetry of spin-up bands and creates a difference in the corresponding gap sizes as shown in Fig. 4C. 4. By further increasing *D*, the K'↑ band gap is inverted, and so is its valley Chern number. Therefore, K↑ and K'↑ bands both contribute $C = -\frac{5}{2}$ that add up to *C* = -5 (or +5 for the state that is the time-reversal version of the illustrated one), corresponding to the QAH state observed. 5. By increasing *D* even further, the K↑ bands also invert, which flips the sign of its valley Chern number and leads to the LP state with zero total Chern number. 6. In the realistic band structure of rhombohedral pentalayer graphene as shown in Fig. 4E (see (*45*)), there is an overlap in the energy range of the conduction and valence bands. Adding this factor into the model would create two semi-metallic regions during phase transitions as observed, but would not change the picture qualitatively. We note that the semi-metallic transition regions could also possibly arise from entropy-driven proliferation of domain walls (*47*), which is beyond the scope of this work.

The mechanism of realizing QAHE in pentalayer graphene is similar to that of the Haldane model for the following reasons: 1. The essence of both mechanisms is that both valleys contribute the same Chern numbers when they are oppositely gapped, which leads to QAHE at charge neutrality of graphene; 2. The Ising SOC term in our model, for the spin bands of smaller gaps, is equivalent to the valley-contrasting mass term in the Haldane model; 3. They both have potential energy difference $\Delta_{AB}$ as a tuning parameter, which is between the sublattices A and B in monolayer graphene in the Haldane model, and between A sublattice in the top layer and B sublattice in the bottom layer in pentalayer graphene. Because of these similarities, we can construct a schematic phase diagram similar to that developed by Haldane. As shown in Fig. 4F, $\Delta_{AB}$ is the *D*-induced potential difference between sublattices as in the Haldane model, whereas $\Delta_{SOC}$ serves as the valley-contrasting mass induced by the phase $\phi$ in the Haldane model. In the triangular region bounded by the semi-metallic phases and $\Delta_{SOC} = \Delta_{LAF}$, pentalayer graphene is in the QAH state. The scenario of $\Delta_{SOC} > \Delta_{LAF}$ is beyond the scope of this work.

**Discussion and outlook**

The QAHE in our system is distinct from those in experimentally reported QAHE systems. Different from magnetic topological insulators, the time-reversal-symmetry in our system is broken spontaneously without the ordering of magnetic elements. In addition, the large Chern number in pentalayer graphene originates from the large pseudo-spin winding number around the Dirac point of graphene (*18, 48, 49*), as an intrinsic property of one layer of the heterostructure. This is different from increasing the Chern number by stacking (effectively decoupled) multiple periods of an MBE-grown topological insulator superlattice (*15*). In 2D moiré superlattices where QAHE was observed (twisted bilayer graphene aligned to hBN (*8*) and twisted TMD (*9, 10, 14*)), the Chern number is contributed by one principal valley owing to the selective population of one isospin band. In contrast, the population in the K and K' valleys and their contributions to the Chern number are both equal in our system. As a result, the QAHE happens at charge neutrality in pentalayer graphene rather than at non-zero charge densities and filling factors as in 2D moiré superlattices.

Our pentalayer graphene/$WS_2$ realizes QAHE in a relatively less-demanding material system, as it is built upon crystalline 2D materials without special twist angles between adjacent layers. This simple structure avoids the inhomogeneity induced by the variation of twist angles in moiré superlattices. The large Chern number is determined by the layer number (*18*, *49*) and could possibly be increased even more by using thicker rhombohedral graphene flakes. In addition, our discovery of QAHE with an odd Chern number has an immediate implication for engineering chiral Majorana edge states towards topological quantum computation: 1. The fact that $WS_2$ is needed on only one side of graphene makes it possible to proximitize the system with a superconductor on the other side as proposed previously (*50*, *51*); 2. Superconductivity has been demonstrated in doped bilayer and rhombohedral trilayer graphene (*21*, *22*, *38*, *41*) and could also possibly exist in thicker rhombohedral graphene layers — implying that a lateral junction between QAHE and the time-reversal-symmetry-broken-superconductivity regions could be potentially engineered to realize chiral Majorana edge states. Our results call for further experiments and theories to understand and explore the intricate interplay between electron correlations, proximity effects and band topologies in 2D material heterostructures.

**Acknowledgments:** We acknowledge helpful discussions with E. Berg, A. Stern, T. Senthil, P. Jarillo-Herrero, K.F. Mak and F. Wang. We acknowledge S. Nadj-Perge and Y. Zhang for their helpful discussions and for sharing unpublished data. We acknowledge L. Shi and S. Ye for their help in sample fabrication. We acknowledge N. Gedik, B. Ilyas and T. Luo for their help in second harmonic generation measurements.

**Funding:** L.J. acknowledges support from a Sloan Fellowship. Work by T.H., J.Y. and J.S. was supported by NSF grant no. DMR- 2225925. The device fabrication of this work was supported by the STC Center for Integrated Quantum Materials, NSF grant no. DMR-1231319 and was carried out at the Harvard Center for Nanoscale Systems and MIT.Nano. Part of the device fabrication was supported by USD(R&E) under contract no. FA8702-15-D-0001. K.W. and T.T. acknowledge support from the JSPS KAKENHI (Grant Numbers 20H00354, 21H05233 and 23H02052) and World Premier International Research Center Initiative (WPI), MEXT, Japan. L.F. was supported by the STC Center for Integrated Quantum Materials (CIQM) under NSF award no. DMR-1231319. C.Y. and F.Z. was supported by NSF under Grants no. DMR-1945351, no. DMR-2105139, and no. DMR-2324033; they acknowledge the Texas Advanced Computing Center (TACC) for providing resources that have contributed to the research results reported in this work.

**Author contributions**: L.J. supervised the project. T.H. and Z.L., performed the DC magneto-transport measurement. T.H. and Y.Y. fabricated the devices. J.Y., J.S., Z.L. and T.H. helped with installing and testing the dilution refrigerator. K.W. and T.T. grew hBN crystals. C.Y. and F.Z. performed the theoretical calculations. L.F. and F.Z. contributed to data analysis. All authors discussed the results and wrote the paper.

**Competing interests**: The authors declare no competing financial interests.

**Data and materials availability**: All the data in the main text and supplementary materials, as well as the code to calculate the band structures, can be obtained at Harvard Dataverse (*52*).

**Supplementary materials:**

Materials and Methods

Supplementary Text

Figs. S1 to S11

References (*53-55*)

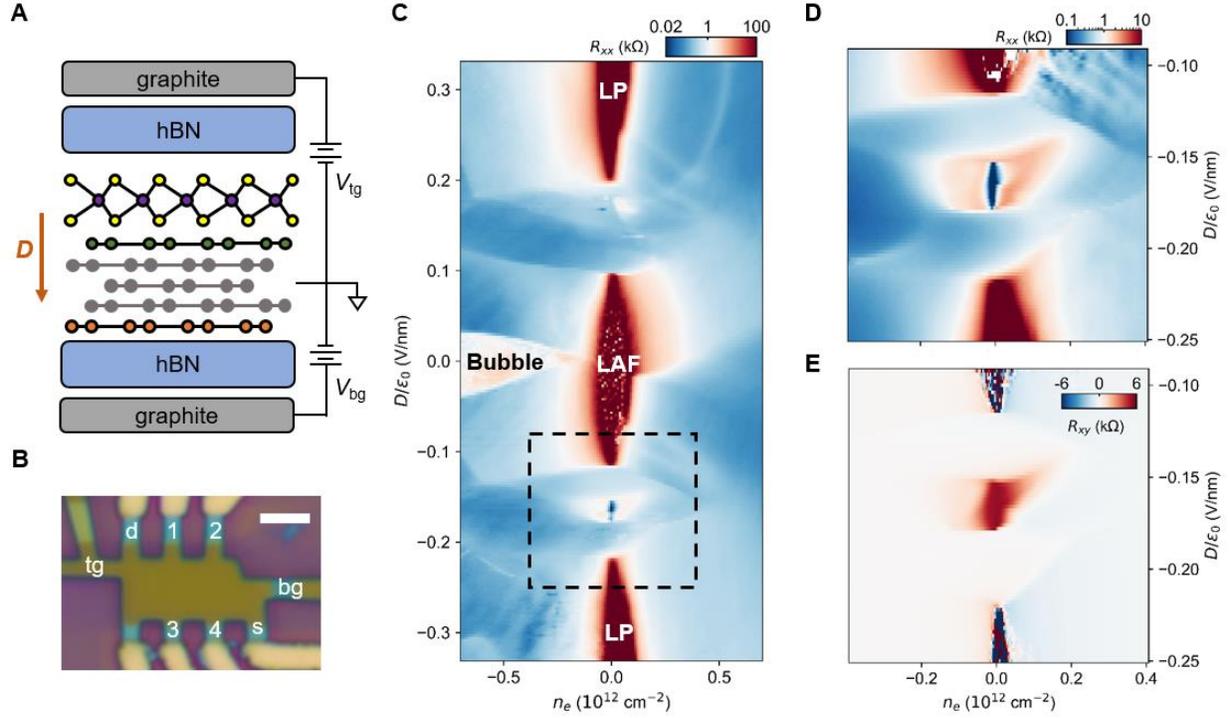

**Fig. 1. Device configuration and phase diagram of rhombohedral pentalayer graphene/monolayer WS$_2$ Device.** (**A**) Schematic of the device configuration, showing a monolayer WS$_2$ neighboring the top layer of rhombohedral graphene. (**B**) Optical micrograph of the device. In the measurement, electrodes labeled as 's' and 'd' are used to pass a current, whereas 1&2 (or 3&4) are used for measuring $R_{xx}$, and 1&3 (or 2&4) are used for measuring $R_{xy}$. Scale bar: 3 μm. (**C**) $R_{xx}$ as a function of $n_e$ and $D$ at $B = 0$ T, featuring LAF, LP and the Bubble phases, similar to rhombohedral pentalayer graphene without WS$_2$. (**D-E**) Zoomed-in maps of symmetrized $R_{xx}$ and anti-symmetrized $R_{xy}$ corresponding to the dashed box in (C) measured at $B = \pm 0.1$T. In the vicinity of $n_e = 0$ cm$^{-2}$ and $D/\varepsilon_0 = -0.165$ V/nm, $R_{xx}$ shows values below 100 Ω and $R_{xy}$ shows values of ~ 5 kΩ, the latter corresponding to a large anomalous Hall resistance. All measurements were taken at the base temperature.

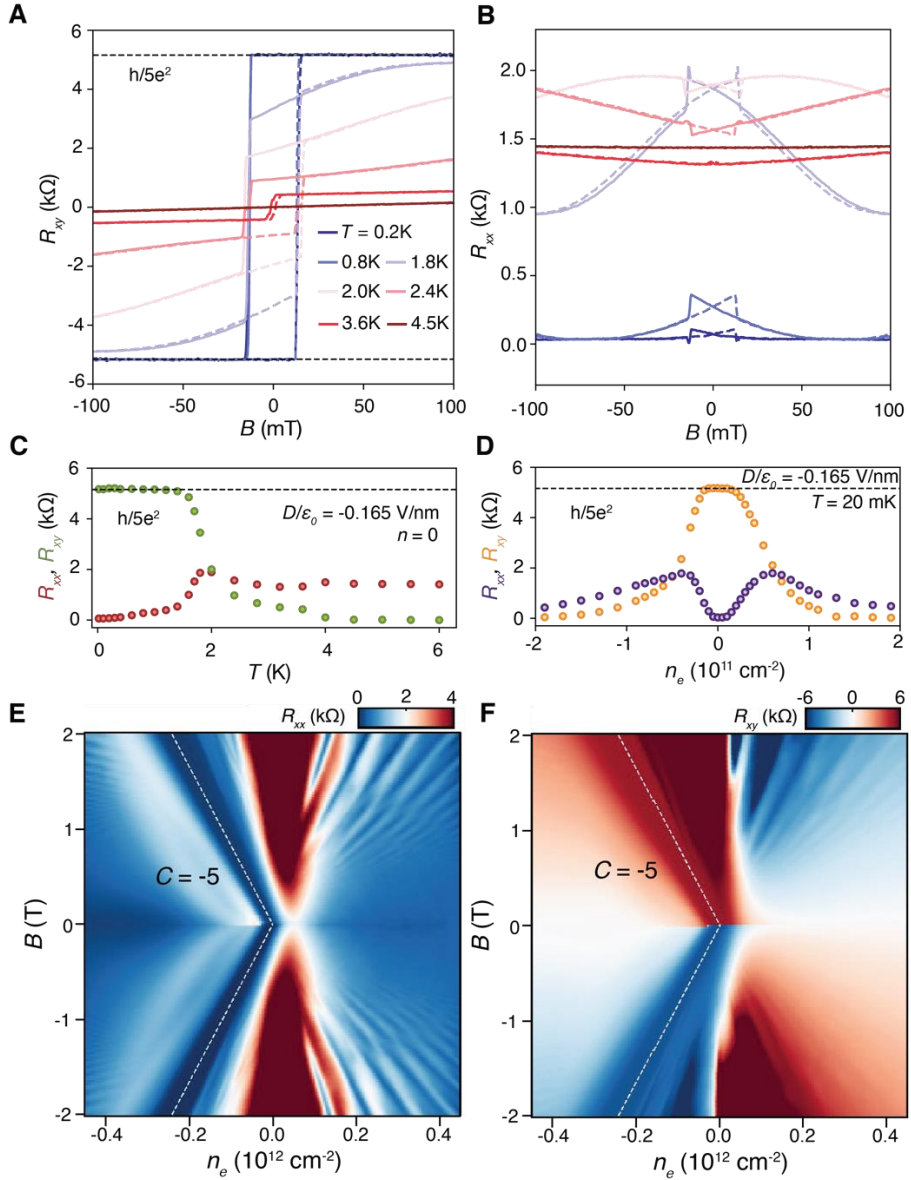

**Fig. 2. Quantum anomalous Hall effect.** (**A-B**) Magnetic hysteresis scan of $R_{xy}$ and $R_{xx}$ at $n_e = 0$ cm$^{-2}$ and $D/\varepsilon_0 = -0.165$ V/nm and $T = 0.2$-$4.5$ K. Solid (dashed) lines correspond to scanning $B$ from positive (negative) values to negative (positive) values. At 0.2 K, $R_{xy}$ is quantized at $\pm 5.16$ k$\Omega$ which corresponds to a Chern number $C = \pm 5$, and $R_{xx}$ shows a value $< 100$ $\Omega$ at $B = 0$ mT. (**C**)

Temperature dependences of zero-magnetic-field $R_{xx}$ and $R_{xy}$ extracted from (A) & (B), showing that $R_{xy}$ reaches the quantized value at ~1.5 K. (**D**) Zero-magnetic-field $R_{xx}$ and $R_{xy}$ as a function of $n_e$ at $D/\varepsilon_0$ = -0.165 V/nm and $T$ = 20 mK, featuring quantized $R_{xy}$ in a plateau of width ~$0.5*10^{11}$ cm$^{-2}$. (**E-F**) Landau-fan maps of $R_{xx}$ and $R_{xy}$ at $D/\varepsilon_0$ = -0.165 V/nm and the base temperature. The QAH state can be seen as wide plateaus in both maps, the slope of which agrees well with that of the dashed lines (corresponding to $C = \pm 5$, as determined by the Streda's formula). At above ~0.5 T, features associated with integer quantum Hall states start to appear for $n_e > 0$ cm$^{-2}$.

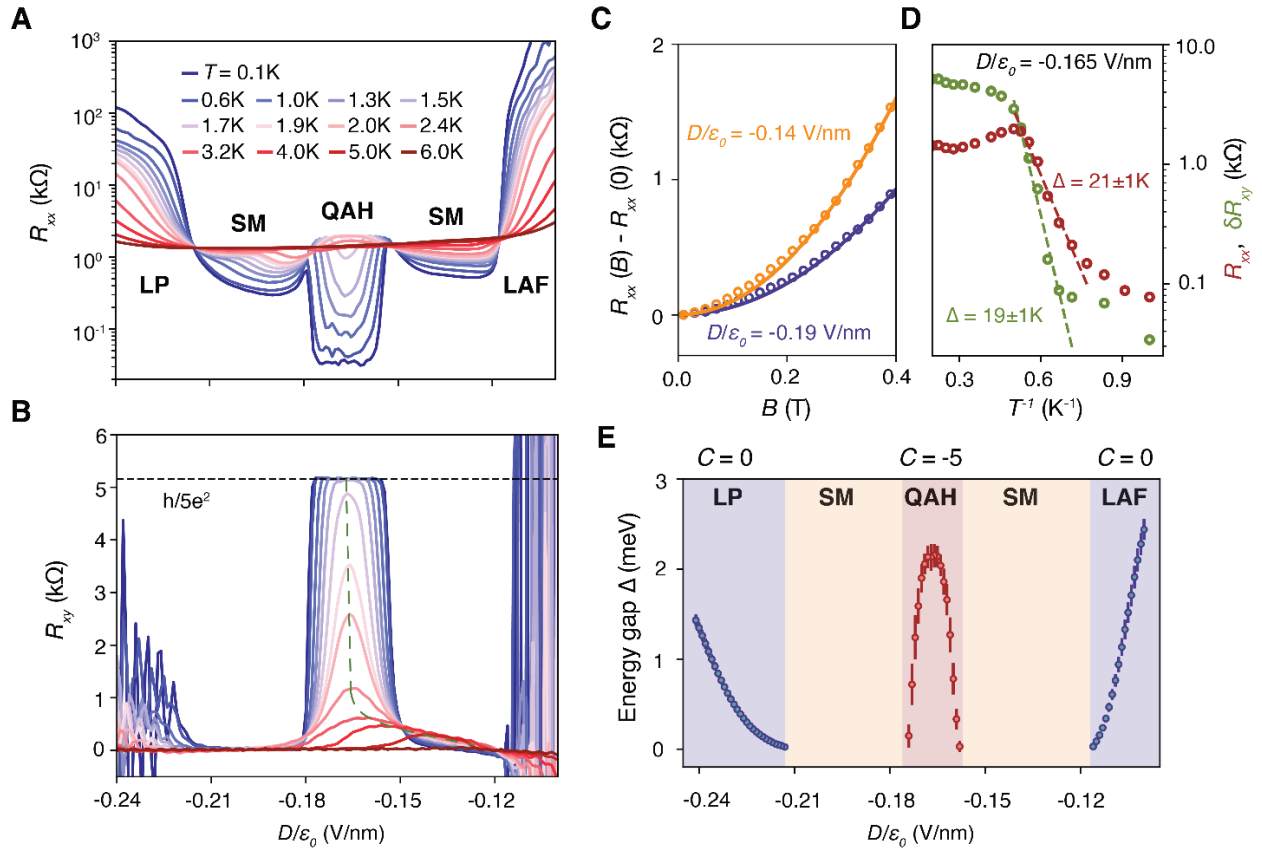

**Fig. 3. Topological phase transitions Driven by *D*.** (**A-B**) $R_{xx}$ and $R_{xy}$ as a function of *D* at $n_e$ = 0 cm$^{-2}$, when the temperature is tuned from 0.1 to 6 K. Three distinct regions can be identified as topologically trivial insulating (including LAF and LP), topologically non-trivial (QAH), and metallic/semi-metallic (SM). $R_{xx}$ and $R_{xy}$ are symmetrized and anti-symmetrized at $B = \pm0.05$ T, respectively. (**C**) Magneto-resistance at $D/\varepsilon_0 = -0.14$ and $-0.19$ V/nm at $n_e = 0$ cm$^{-2}$, which can be fitted with a $B^2$ dependence as typical for semi-metals but not metals. The data were taken at the base temperature. (**D**) Temperature dependences of $R_{xx}$ and $\delta R_{xy} = \frac{h}{5e^2} - R_{xy}$ at $D/\varepsilon_0 = -0.165$ V/nm and $n_e = 0$ cm$^{-2}$. Before the semi-metal phase shifts into this range of $D$ and dominates the behavior, both $R_{xx}$ and $\delta R_{xy}$ follow a thermal activation model with gap sizes of 21 K and 19 K respectively. $R_{xx}$ and $R_{xy}$ are symmetrized and anti-symmetrized at $B = \pm0.05$ T, respectively. (**E**) Summary of the band gap sizes as a function of $D$. The data points in the LP, LAF and QAH regions are extracted from $R_{xx}$ fittings with a thermal activation model with the uncertainty labelled as vertical bars (see Fig.S4 (*45*)).

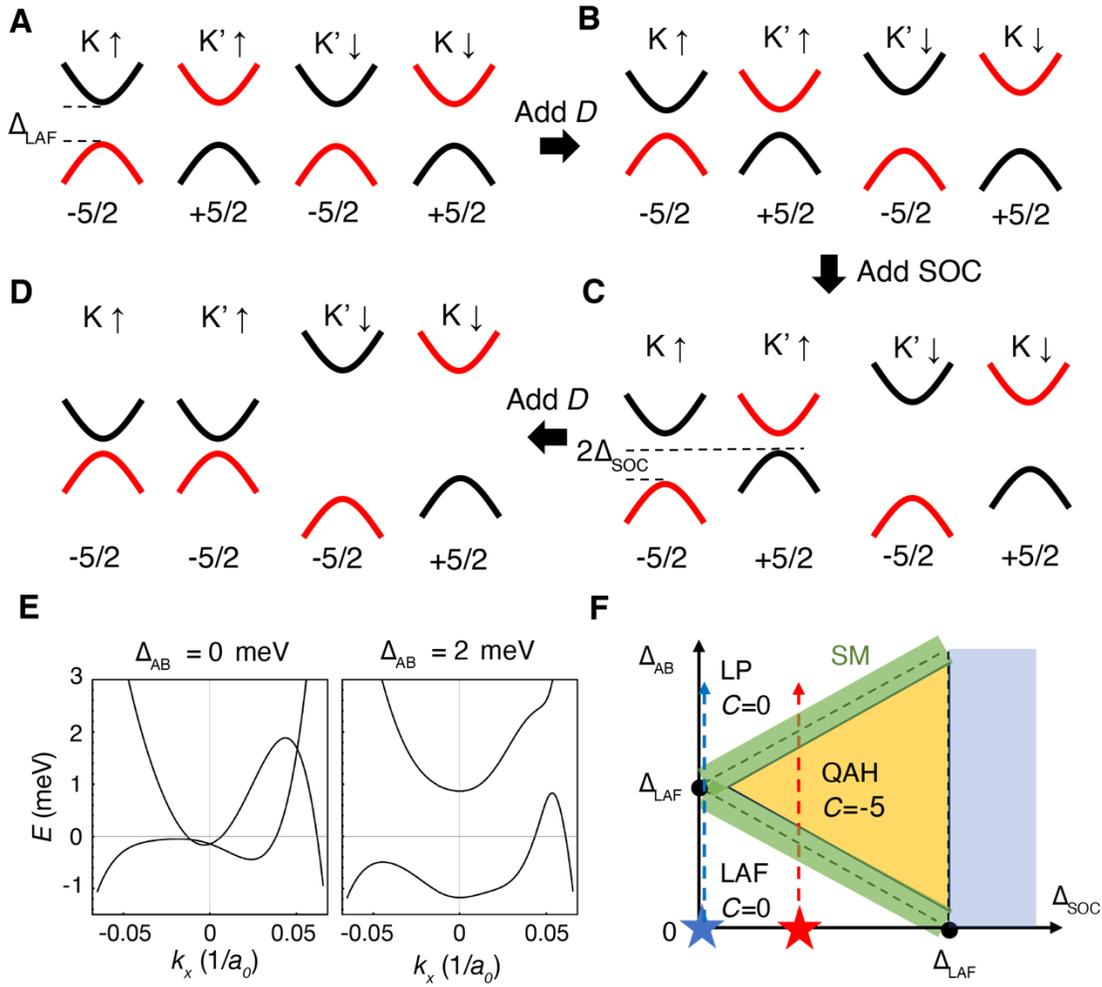

**Fig. 4. A model for the topological phase transitions triggered by electron correlation, displacement field, and spin-orbit-coupling.** (**A**) Illustration of the band structure of pentalayer graphene at zero $D$ and zero SOC from WS$_2$. The details of the band structures such as trigonal warping and electron-hole asymmetry are neglected. Red and black colors of each isospin band structure correspond to -5/2 and +5/2 valley Chern numbers of that band, respectively. The valley Chern number of the valence band is shown below the band structure. In this LAF state (with a spontaneous band gap $\Delta_{LAF}$), spin-up and spin-down states polarize to the top and bottom layers respectively. (**B**) Band structures when a non-zero $D$ is applied, which reduces the gap of spin-up bands and enlarges the gap of spin-down bands. The time-reversal-symmetry is spontaneously

broken and the model is simplified to a spinless model if only low-energy bands are considered. (**C**) Low-energy band structures when WS$_2$-induced Ising SOC is introduced, where the valley-symmetry is explicitly broken by $\Delta_{SOC}$. (**D**) By fine-tuning *D*, the gap, the corresponding sign of the valley Chern number, and the layer polarization of states near the band edges of K' are inverted. This results in a total Chern number *C*=-5 and the QAH state. (**E**) The realistic band structures of rhombohedral pentalayer graphene at interlayer potential $\Delta_{AB} = 0$, 2 meV (*45*). The former shows the overlap between conduction and valence bands which induces the semi-metallic regions in (F); a clean global gap can be created by $\Delta_{AB} > 2$ meV. (**F**) Schematic of the phase diagram as a function of displacement field-induced interlayer potential difference $\Delta_{AB}$ and WS$_2$-induced SOC strength $\Delta_{SOC}$. The stars and dotted arrows label the phase transitions in the pentalayer without WS$_2$ (blue) and with WS$_2$ (red).

## Materials and Methods

### Device fabrication

The pentalayer graphene, monolayer WS$_2$ and hBN flakes were prepared by mechanical exfoliation onto SiO$_2$/Si substrates. The rhombohedral domains of pentalayer graphene were identified using near-field infrared microscopy (*53*), confirmed with Raman spectroscopy and isolated by cutting with a Bruker atomic force microscope (AFM) (*54*). The van der Waals heterostructure was made following a dry transfer procedure. We picked up the top hBN, graphite, middle hBN, the WS$_2$, and the pentalayer graphene using polypropylene carbonate (PPC) film and landed it on a prepared bottom stack consisting of an hBN and graphite bottom gate. We intentionally misaligned the straight edges of the graphene and hBN flakes to avoid the effects of the moiré superlattices. The twist angle between WS$_2$ and graphene is estimated to be ~7 degrees based on the straight edges of the WS$_2$ and graphene flake (Fig. S1D&E). The device was then etched into a Hall bar structure using standard e-beam lithography (EBL, Elionix F125) and reactive-ion etching (RIE). We deposited Cr/Au for electrical connections to the source, drain and gate electrodes. The device shows good homogeneity as the QAHE can be seen by using different pairs of transverse voltage probes, as shown in Fig. S1A-C.

### Transport measurement

The device was measured in a Bluefors LD250 dilution refrigerator with an electronic temperature of around 100 mK. Stanford Research Systems SR830 lock-in amplifiers were used to measure the longitudinal and Hall resistance $R_{xx}$ and $R_{xy}$ with an AC voltage bias 80 uV at a frequency at 17.77 Hz. Keithley 2400 source-meters were used to apply top and bottom gate voltages. Top-gate voltage ($V_t$) and bottom-gate voltage ($V_b$) are swept to adjust doping density $n_e = (C_t V_t + C_b V_b)/e$ and displacement field $D = (C_t V_t - C_b V_b)/2$, where $C_t$ and $C_b$ are top-gate and bottom-gate capacitance per area calculated from the Landau fan diagram.

### Effective tight-binding model informed by first-principles calculations

For pentalayer graphene with the rhombohedral stacking, an effective tight-binding model Hamiltonian of Slonczewski-Weiss-McClure type can be written as (*55*)

$$H = \begin{pmatrix} u/2+\delta & v_0\pi^\dagger & v_4\pi^\dagger & v_3\pi & 0 & \gamma_2/2 & 0 & 0 & 0 & 0 \\ v_0\pi & u/2 & \gamma_1 & v_4\pi^\dagger & 0 & 0 & 0 & 0 & 0 & 0 \\ v_4\pi & \gamma_1 & u/4 & v_0\pi^\dagger & v_4\pi^\dagger & v_3\pi & 0 & \gamma_2/2 & 0 & 0 \\ v_3\pi^\dagger & v_4\pi & v_0\pi & u/4 & \gamma_1 & v_4\pi^\dagger & 0 & 0 & 0 & 0 \\ 0 & 0 & v_4\pi & \gamma_1 & 0 & v_0\pi^\dagger & v_4\pi^\dagger & v_3\pi & 0 & \gamma_2/2 \\ \gamma_2/2 & 0 & v_3\pi^\dagger & v_4\pi & v_0\pi & 0 & \gamma_1 & v_4\pi^\dagger & 0 & 0 \\ 0 & 0 & 0 & 0 & v_4\pi & \gamma_1 & -u/4 & v_0\pi^\dagger & v_4\pi^\dagger & v_3\pi \\ 0 & 0 & \gamma_2/2 & 0 & v_3\pi^\dagger & v_4\pi & v_0\pi & -u/4 & \gamma_1 & v_4\pi^\dagger \\ 0 & 0 & 0 & 0 & 0 & 0 & v_4\pi & \gamma_1 & -u/2 & v_0\pi^\dagger \\ 0 & 0 & 0 & 0 & \gamma_2/2 & 0 & v_3\pi^\dagger & v_4\pi & v_0\pi & -u/2+\delta \end{pmatrix}$$

in the basis of (A$_1$, B$_1$, A$_2$, B$_2$, A$_3$, B$_3$, A$_4$, B$_4$, A$_5$, B$_5$). Here $v_i = \sqrt{3}a_0\gamma_i/2\hbar$ with $a_0 = 0.246$ nm. We have accurately fit the model parameters to match the band structure obtained from our first-principles calculations that will be reported in *Band Theory of Rhombohedral Graphene*; for pentalayer $\gamma_0 = 3.16$ eV, $\gamma_1 = 0.435$ eV, $\gamma_2 = -0.0185$ eV, $\gamma_3 = -0.322$ eV, $\gamma_4 = -0.0675$ eV, and $\delta = -0.147$ meV. An interlayer electric displacement field introduces a potential difference between the top and bottom layers, denoted by $u$ in the Hamiltonian, and here we assume the gradient of the induced potential is uniform across the five graphene layers. By using $u = 0$ and 2 meV we can obtain Fig. 4E in the main text. A more realistic modeling of the large Chern number quantum anomalous Hall states that incorporates the $WS_2$ induced spin-orbit couplings and extra screening, the electron-electron interactions of both long-range and short-range, and the trigonal warping effect will be reported in *Proximitized Rhombohedral Graphene: Quantum Anomalous Hall Factory*.

**Supplementary Text**

Behaviors of $R_{xx}$ and $R_{xy}$ at the positive $D$ side

Fig. S2 shows the longitudinal and transverse resistances at the positive $D$ side. No signature of QAHE has been seen down to the base temperature. However, there is a small anomalous Hall signal in the semi-metallic region between the LAF and LP regions (Fig. S8). Upon applying a vertical magnetic field, Chern insulators with $C = \pm 5$ can be induced at a magnetic field threshold value of ~0.2 T. This value is significantly lower than that in pentalayer rhombohedral graphene without $WS_2$. This observation indicates that SOC in general drives the system towards QAH states at both positive and negative $D$s, but there might be a subtle difference between the two sides. This asymmetry about $D$ is likely related to the fact that $WS_2$ is neighboring the top layer of rhombohedral graphene. A complete understanding requires theoretical calculations that include the details of SOC, the accurate hopping parameters in pentalayer graphene, and proper considerations of many-body effects —beyond the scope of this experimental work.

Gap extraction from thermal activation

We obtain the transport gap for the LP, LAF and QAH states from thermal activation behavior of $\sigma_{xx}$, as in ref (*8*). For the topological trivial insulating states LP and LAF, we relate $\sigma_{xx}$ to the measured $R_{xx}$ by $\sigma_{xx} = 1/R_{xx} = \sigma_0 e^{-\Delta/2T}$. Fig. S4 shows examples of $R_{xx}$ at LAF and LP states and the gap size summary in presented in Fig. 3E. For the QAH state, we model $\sigma_{xx}$ as thermally activated and $\sigma_{xy}$ as roughly constant. When $\sigma_{xx}$ is small, we have $\rho_{xx} = \sigma_{xx}/(\sigma_{xx}^2 + \sigma_{xy}^2) \sim \sigma_{xx}/\sigma_{xy}^2$ and $\rho_{xy} = \sigma_{xy}/(\sigma_{xx}^2 + \sigma_{xy}^2) \sim 1/\sigma_{xy} - \sigma_{xx}^2/\sigma_{xy}^4$. As a result, we can extract the gap by fitting $R_{xx} \propto \sigma_{xx} = R_0 e^{-\Delta/2T}$ and $\delta R_{xy} = h/5e^2 - R_{xy} \propto \sigma_{xx}^2 = R_1 e^{-\Delta/T}$. Examples of $R_{xx}$ and $\delta R_{xy}$ are shown in Fig. 3D and the summary of gap extracted from $R_{xx}$ and $\delta R_{xy}$ is presented in Fig. S4B and the two sets of gaps agree reasonably well.

The faint feature at low magnetic fields

We note that the faint state neighboring the $C = \pm 5$ state at low magnet fields is not a well-defined quantum-Hall state. Fig. S5A shows the $n_e$ - $D$ map of $R_{xx}$ at 0.4T where the additional state neighboring the Chern state shows up in Fig. 2E of the main text. The additional state in Fig. 2E corresponds to the faint tilted feature pointed by the black arrow. This state is not a well-defined quantum-Hall state since it does not stay at a fixed density in the $n_e$ - $D$ map.

We further measured the Landau fan at a smaller and larger $D$ compared to Fig. 2E, as shown in Fig. S5B-D (we reproduced Fig. 2E in Fig. S5C for comparison). The choice of $D$ is indicated by the three colored dots in Fig. S5A, and they all cut into the tilted feature. Although the features connect smoothly with each other in the $n_e$ - $D$ map, their magnetic-field dependences are different. As shown in Fig. S5B-D, the $C = \pm 5$ states show the same slope and project to zero density at all three $D$'s. However, the additional features have different slopes and project to different densities at zero magnetic field. Additionally, the state is very faint and does not develop better with increasing the magnetic field. So we tend to believe it is not a quantum anomalous Hall state.

The state on the other side of $D$ looks more likely to be a not-well-developed quantum Hall state with a Chern number $\pm 7$, as shown in Fig. S6. It projects to zero density instead of a finite density.

The quantum oscillation at high hole doping

The 'quantum oscillations' at larger hole doping in Fig. 2E are due to Landau level crossings, originating from the multiple Fermi surfaces with different sizes. Such effects are typical in rhombohedral trilayer and bilayer graphene systems (*21–23*, *41*). To clarify on this, we have performed more measurements with extended ranges of $n_e$ and $B$ compared to Fig. 2E and shown them in Fig. S7. Fig. S7A shows the Landau fan covering these states on the negative $D$ side. At magnetic fields larger than 1.7 T, we can clearly see short segments within the dark stripes, where the $R_{xx}$ is almost zero. These segment features correspond to when the Fermi level is inside a quantum Hall gap between two adjacent Landau levels. All such segments extend to the charge neutral point at $B = 0$; see the white dashed lines for examples. When tuning $B$ and/or $n$, these segments are intercepted by some white stripes (corresponding to large $R_{xx}$ when the Fermi level is inside a Landau level) which arise from the Landau level crossings at the Fermi level. At lower magnetic fields, the segments are shorter and more difficult to resolve, resulting in visually misleading dark and white stripes that resemble quantum Hall states.

We have also performed the same measurement for the positive $D$ side. The behavior is similar to the negative D side, as shown in Fig. 7B. Fig S2C does not cover a large enough density range, so such features were not visible.

Anomalous Hall effect on positive $D$ side.

Fig. S8 shows $R_{xy}$ vs $B$ sweeps on the positive $D$ side. Although the anomalous Hall signal is much smaller, the magnetic hysteresis and anomalous Hall $R_{xy}$ are clear. We believe the anomalous Hall effect on the positive $D$ side is from a semimetal with Berry curvature.

In our toy model, the induced SOC cooperates with the e-e interaction to create opposite energy gaps at K and K' valleys for one spin species, very similar to that in the Haldane model. However, due to the intrinsic band overlap, SOC or e-e interaction effect needs to be big enough to open a global bandgap for the QAH insulator. If SOC or e-e interaction effect is not big enough compared to the band overlap (which might be complicated by the possible screening effect in graphene by $WS_2$, due to the asymmetric sample geometry vertically), the state at intermediate $D$ between LAF and LP state could be a semimetal with direct bandgap everywhere but no global bandgap across the whole Brillouin Zone. In this case, there are both electrons and holes in the system, with opposite signs of Berry curvature (see Fig. S9A for a more realistic calculation). Their contribution to the anomalous Hall and orbital magnetization would add up. Since the low-energy K and K' valleys have the same sign of Berry curvature in this semimetal, the total anomalous Hall and orbital magnetization do not need to be zero.

The magnetoresistance (MR, defined as $(R_{xx}(B)-R_{xx}(0))/R_{xx}(0)$) follows a $B^2$ behavior, indicating the state at the positive $D$ side likely to be an electron-hole compensated semimetal (Fig. S9B).

Evidence for the presence of SOC

The Ising SOC causes an energy shift of bands with different spins in a given valley (*31*, *32*, *36*, *38*, *41*) breaking the isospin degeneracy, as shown in Fig. S10A. This results in mismatched Fermi surface areas of the favored and disfavored bands, which can be detected with the quantum oscillation (QO) measurement (Fig. S10B) (*22*, *23*, *38*, *41*). We plotted in Fig. S10C the Fourier transform of $R_{xx}(1/B)$ as a function of density $n_e$ and normalized frequency, $f_v = f_B/(\phi_0 n_e)$, where $f_B$ is the QO frequency and $\phi_0 = h/e$ is the magnetic flux quantum. $f_v$ can be interpreted as the fractional share of the total electrons enclosed by a given Fermi contour. Peaks around $f_v = ¼$ corresponds to the four Fermi surfaces from the four isospins. The splitting of the peak (also visible at $f_v = ½$), especially at low density, indicates the four Fermi surfaces have two different sizes and we attribute this to the presence of SOC. Such methodology has been adopted in other works such as in Bernal bilayer graphene/WSe$_2$ (*38, 41*).

Our device with WS$_2$ (#1) also connects to another one (#2) that shares the same graphene but without WS$_2$ on the top, as shown in Fig. S11A. Device #2 thus serves as a reference. We did not observe QAHE in device #2 down to the base temperature. We performed QO measurement at the same *D* field (Fig. S11C). Without SOC, Fermi surface of the four isospins should have the same area (Fig. S11B), and it is indeed the case in our Fourier transform data (Fig. S11D). A single peak is visible at $f_v$ = ¼ and $f_v$ = ½, indicating the absence of SOC. We also note that in device #2 the QO is clear at a magnetic field as low as 0.5 T, similar to that in device #1. This indicates that the two devices have similar quality.

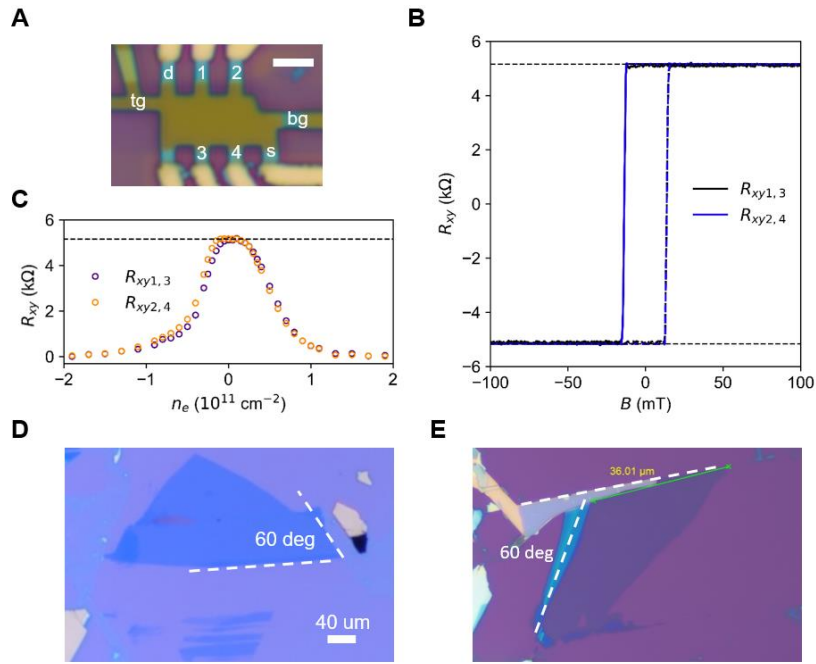

**Fig. S1. Homogeneity of device.** (**A**) Optical micrograph of the device. Scale bar: 3 μm. (**B**) Magnetic hysteresis of the anomalous Hall signals measured with two different pairs of transverse probes. Both signals show quantization at $\frac{h}{5e^2}$ (the dashed line). Solid and dashed lines correspond to forward and backward scans respectively. (**C**) Anomalous Hall signals measured with two different pairs of transverse probes as a function of charge density. Both signals show quantization at $\frac{h}{5e^2}$ (the dashed line) and almost overlapping trajectories. (**D-E**) Optical micrograph of the graphene and WS$_2$ flake on the Si/SiO$_2$ substrate. The white dashed lines indicate the straight edges. The relative rotation of the two flakes in the device is around 7 degrees between their straight edges (the bottom straight edge of graphene and the lower left straight edge of WS$_2$).

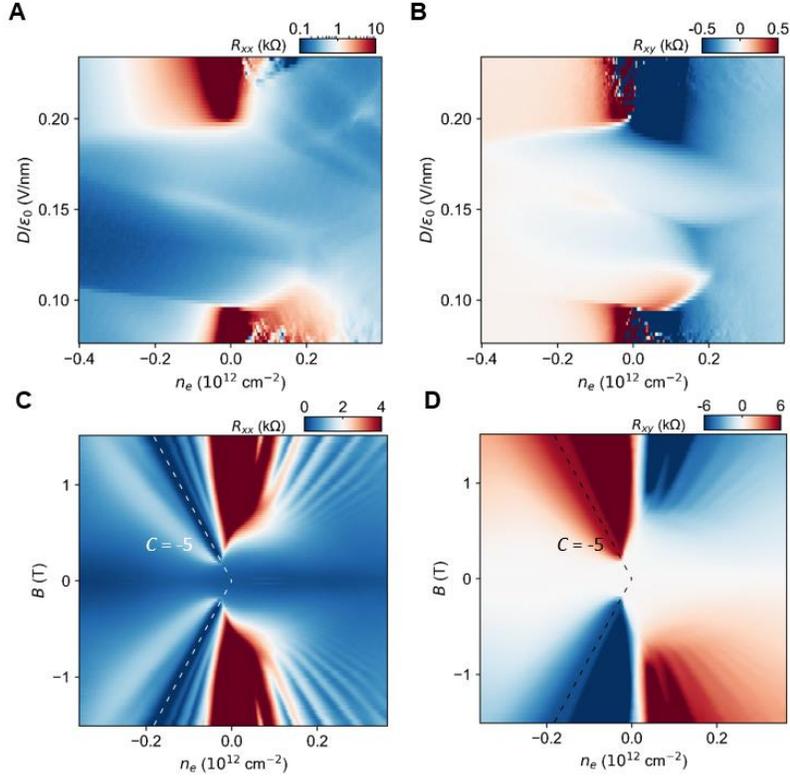

**Fig. S2. Detailed characterizations at the positive D side.** (**A-B**) $R_{xx}$ and $R_{xy}$ as a function of $D$ and $n_e$, showing weak anomalous Hall signals on the order of 50 Ohm and no obvious dip in $R_{xx}$. The maps are generated by symmetrization and anti-symmetrization of maps taken at ±0.1 T. (**C-D**) Landau fan of $R_{xx}$ and $R_{xy}$ at $D/\varepsilon_0 = 0.156$ V/nm, showing Chern insulators with $C=\pm 5$ at a threshold magnetic field of ~0.2 T.

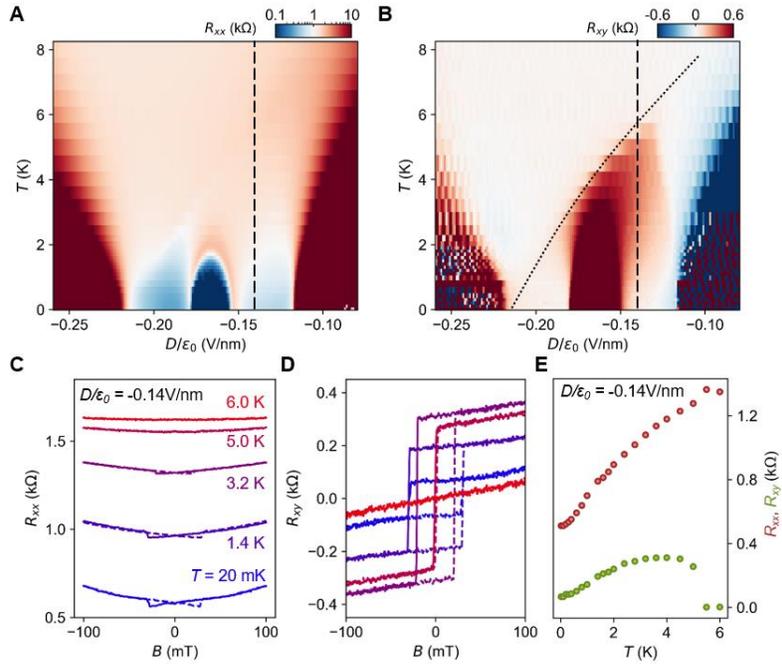

**Fig. S3. Temperature dependence of $R_{xx}$ and $R_{xy}$ in the SM phase.** (**A-B**) $R_{xx}$ and $R_{xy}$ as a function of $D$ and $T$. Clear anomalous Hall signals can be seen in the SM region between QAH and LAF regions at elevated temperatures. The dotted line indicates the boundary of magnetic ordering, and the tilting feature is due to the weakening of LAF as the temperature increases. (**C-D**) Magnetic hysteresis of $R_{xx}$ and $R_{xy}$ at $D/\varepsilon_0$ = -0.14 V/nm at varied temperatures, corresponding to the dashed line in (A&B). (**E**) Temperature dependence of the zero-magnetic-field values of $R_{xx}$ and $R_{xy}$ at $D/\varepsilon_0$ = -0.14 V/nm. We note that the onset temperature of anomalous Hall is larger than that at $D/\varepsilon_0$ = -0.165 V/nm.

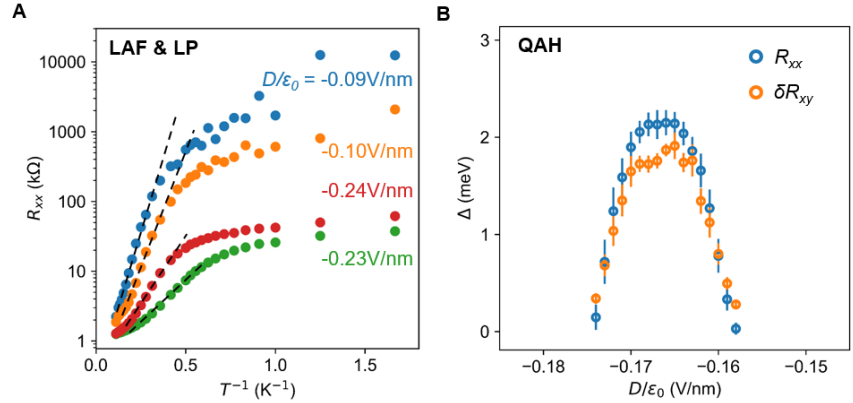

**Fig. S4. Gap extraction from thermal activation.** (**A**) Arrhenius plot of $R_{xx}$ at different $D/\varepsilon_0$. The high-temperature data can be fitted by the thermal activation behavior (dashed line), from which we extract the LAF gap and LP gap as shown in Fig. 3E. (**B**) The summary of the QAH gap extracted from thermal activation fitting of the $R_{xx}$ and $\delta R_{xy} = \frac{h}{5e^2} - R_{xy}$ in Fig. 3D.

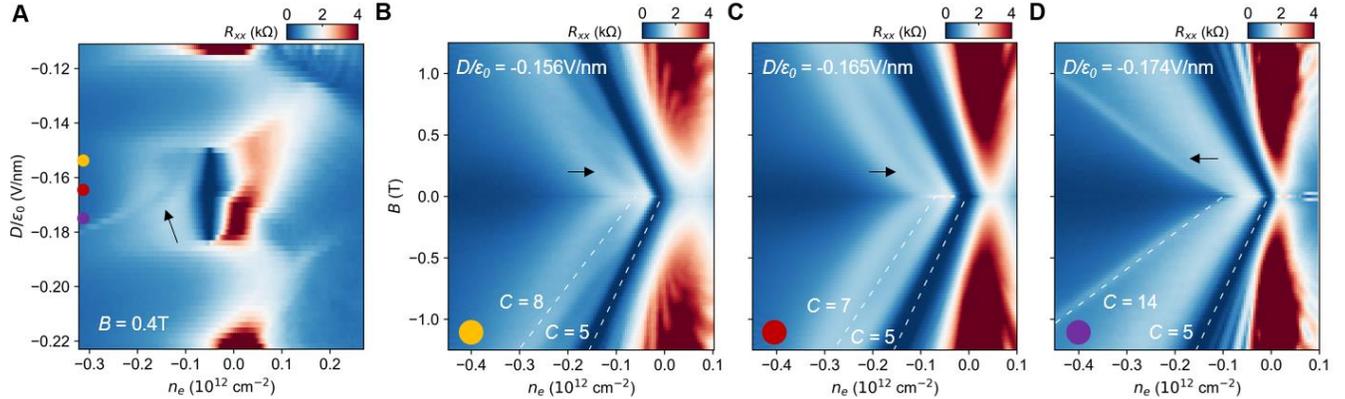

**Fig. S5. The faint feature at low magnetic field on the negative $D$ side.** (**A**) $R_{xx}$ as a function of $D$ and $n_e$, measured at $B$ = 0.4T. The black arrow marks the additional state in Fig. 2E, a dip of $R_{xx}$ on the lower density side of the $C = \pm 5$ states. (**B-D**) Landau fan of $R_{xx}$ at $D/\varepsilon_0$ = -0.156 V/nm, -0.165 V/nm and -0.174 V/nm respectively. The colored dot corresponds to those in Fig. S5A, where they mark the chosen $D/\varepsilon_0$. The $C = \pm 5$ states show the same slope and project to zero density. The additional features (marked by black arrows) have different slopes and project to different densities at zero magnetic field. The white dashed lines are drawn based on Streda's formula for the Chern number specified next to them.

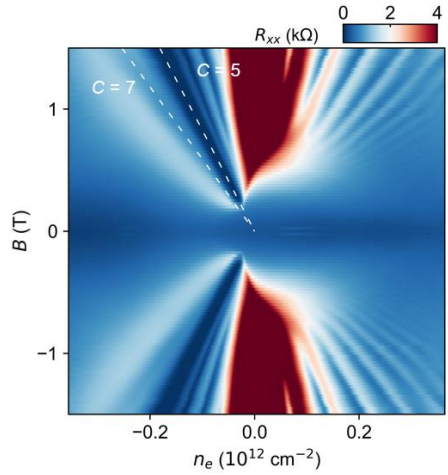

**Fig. S6. The faint feature at low magnetic field on the positive *D* side.** Landau fan of $R_{xx}$ at $D/\varepsilon_0 = 0.156$ V/nm. The faint feature close to the $C = \pm 5$ Chern insulator looks like a not-well-developed quantum Hall state with $C = \pm 7$.

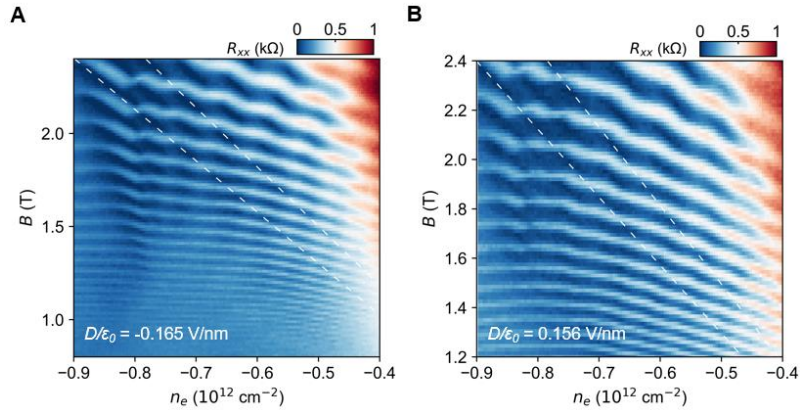

**Fig. S7. Landau fan with extended range of $n_e$ and $B$.** (**A, B**) Landau fan of $R_{xx}$ at $D/\varepsilon_0 = -0.165$ V/nm (A) and 0.156 V/nm (B). The dark stripes correspond to the quantum Hall states and they are only clear when magnetic fields are larger than 1.7 T. These states project to the charge neutral point at $B = 0$ (see the white dashed guidelines for examples).

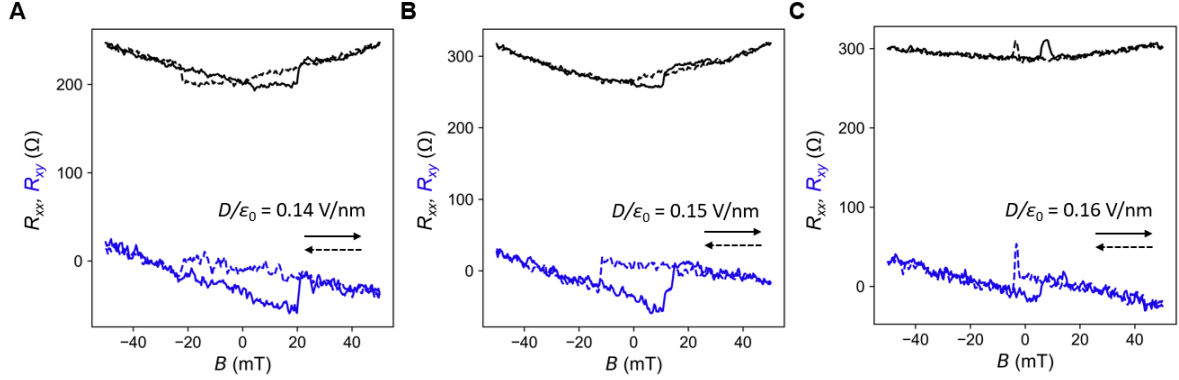

**Fig. S8. Anomalous Hall effect at positive D side.** $R_{xx}$ and $R_{xy}$ at $D/\varepsilon_0 = 0.14$ V/nm (**A**), 0.15 V/nm (**B**), 0.16 V/nm (**C**), and $n_e = 0$ at the base temperature. Solid and dashed lines denote the opposite B field scanning directions.

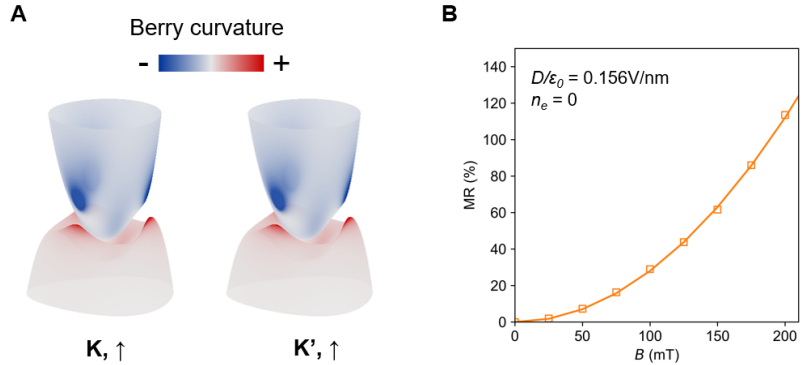

**Fig. S9. Anomalous Hall effect in a semimetal.** (**A**) A tight-binding band structure schematic at around $\Delta_{AB} = \Delta_{LAF}$ where $\Delta_{LAF} < \Delta_{SOC}$. The band structure is color-coded with the Berry curvature. We only plotted the low-energy spin-up bands, because the two high-energy spin-down bands have larger bandgaps and opposite Berry curvatures that cancel each other. The system is a semimetal with direct bandgap everywhere but no global bandgap across the whole Brillouin Zone. The K and K' valleys have the same sign of Berry curvature, while the electron and hole states in the same valley have opposite signs of Berry curvature. This leads to non-zero intrinsic anomalous Hall effect and orbital magnetization. (**B**) The magnetoresistance (MR, defined as $(R_{xx}(B)-R_{xx}(0))/R_{xx}(0)$) at $D/\varepsilon_0 = 0.156$ V/nm and $n_e = 0$ follows a $B^2$ behavior, indicating the state at the positive D side likely to be an electron-hole compensated semimetal.

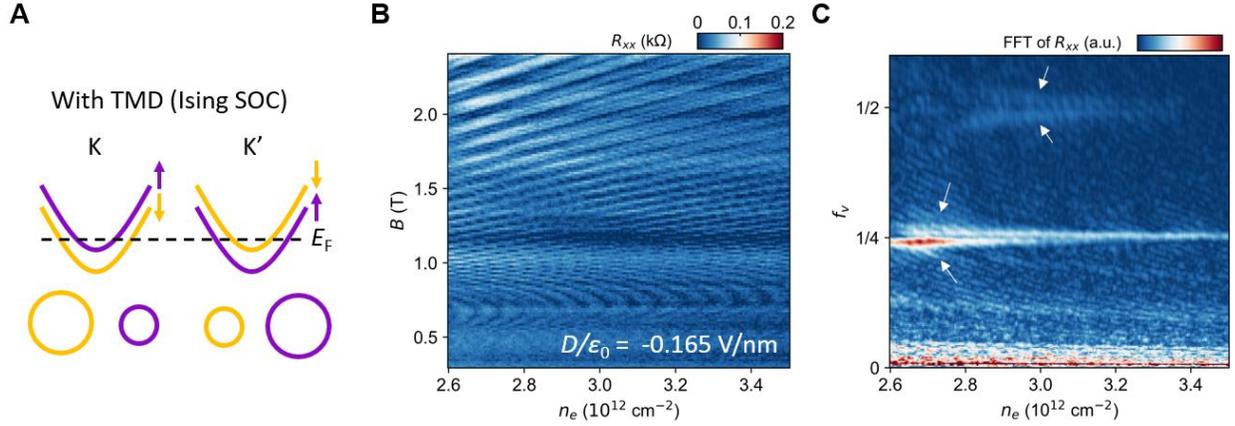

**Fig. S10. Evidence for SOC from quantum oscillations.** (**A**) Schematics of conduction bands of the four isospins with Ising SOC. The SOC lifts the degeneracy of the bands with opposite spins in each valley and they are time-reversal related to the bands in the other valley. The bottom circles (two big and two small) depict the Fermi surfaces of the four isospins. (**B**) $R_{xx}$ as a function of $n_e$ and $B$ at $D/\varepsilon_0 = -0.165$ V/nm at the base temperature. (**C**) We performed Fourier transforms for data in (**B**) at each $n_e$ with respect to $(1/B)$ and plotted them as a function of $f_v = f_B/(\phi_0 n_e)$, where $f_B$ is the quantum oscillation frequency and $\phi_0 = h/e$ is the magnetic flux quantum. $f_v$ can be interpreted as the fractional share of the total electrons enclosed by a given Fermi contour. Peak around $f_v = ¼$ splits into two peaks, especially at low density, indicating the four Fermi surfaces have two sizes and we attribute this to the presence of SOC. Such splitting is also visible in the second harmonic of the quantum oscillation, as shown near $f_v = ½$.

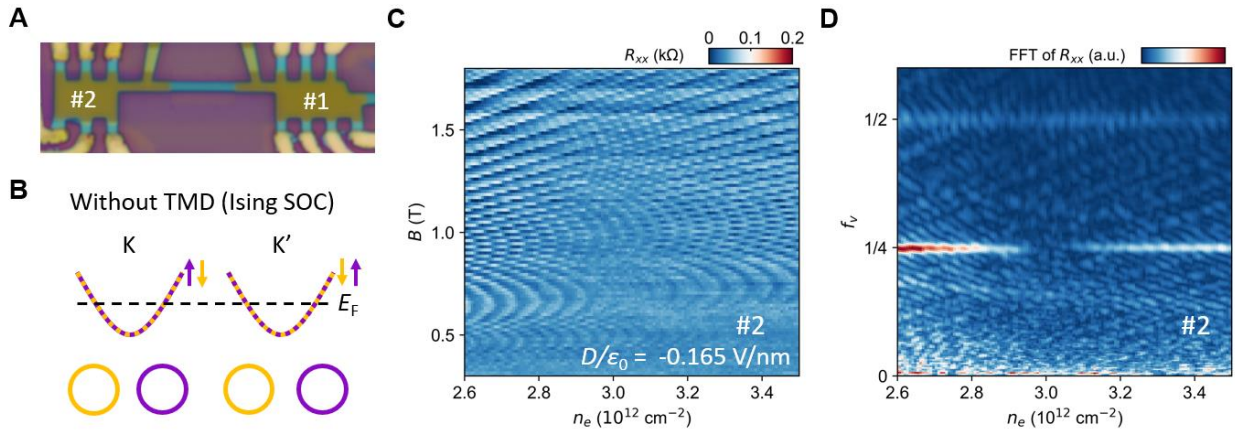

**Fig. S11. 5LG without WS$_2$ as a comparison.** (**A**) Optical image of the devices. #1 is the device discussed in the main text. #2 shares the same graphene with #1 but without WS$_2$ on the top. (**B**) Similar to Fig. S10A, but without TMD. All four isospin bands are degenerate and they have the same Fermi surface area. (**C, D**) Similar to Fig. S10B-C, but in #2. There is only one single peak at $f_v = ¼$, likewise at $f_v = ½$, consistent with the absence of SOC.